%% file: main.tex
\documentclass[sigconf,nonacm]{acmart}

\usepackage{algorithm}
\usepackage{algorithmicx}
\usepackage{bm}
\usepackage{colortbl}
\usepackage{enumitem}
\setlist{topsep=0pt, leftmargin=*}
\usepackage{framed}
\usepackage[utf8]{inputenc}
\usepackage{listings}
\usepackage{multirow}
\usepackage{textcomp}
\usepackage{subfigure}
\usepackage{subcaption}
\usepackage{xspace}    

\def\ouralg{\textsc{FaSTED}\xspace}
\def\tedjoin{\textsc{TED-Join}\xspace}
\def\tedjoinbrute{\textsc{TED-Join-Brute}\xspace}
\def\tedjoinindex{\textsc{TED-Join-Index}\xspace}
\def\gdsjoin{\textsc{GDS-Join}\xspace}
\def\mistic{\textsc{MiSTIC}\xspace}

\def\ptxmma{PTX \texttt{mma}}

\def\tinyds{\textit{Tiny5M}\xspace}
\def\sift{\textit{Sift10M}\xspace}
\def\gist{\textit{Gist1M}\xspace}
\def\cifar{\textit{Cifar60K}\xspace}

\def\synth{\textit{Synth}\xspace}

\definecolor{OliveGreen}{cmyk}{0.64,0,0.95,0.40}
\definecolor{CadetBlue}{cmyk}{0.62,0.57,0.23,0}
\definecolor{lightlightgray}{gray}{0.93}
\definecolor{TFFrameColor}{rgb}{0.0,0.0,0.5} 
\definecolor{TFBackgroundColor}{rgb}{0.9,0.95,1} 
\definecolor{TFTitleColor}{rgb}{1,1,1} 

\input{textResources}

\begin{document}

\title{Fast and Scalable Mixed Precision Euclidean Distance Calculations Using GPU Tensor Cores}

\author{Brian Curless}
\affiliation{%
  \institution{School of Informatics, Computing, and Cyber Systems\\Northern Arizona University}
  \city{Flagstaff}
  \state{AZ}
  \country{USA}
}
\email{bc2497@nau.edu}

\author{Michael Gowanlock}
\affiliation{%
  \institution{School of Informatics, Computing, and Cyber Systems\\Northern Arizona University}
  \city{Flagstaff}
  \state{AZ}
  \country{USA}
}
\email{Michael.Gowanlock@nau.edu}
    



\renewcommand{\shortauthors}{Curless and Gowanlock}

\begin{abstract}
Modern GPUs are equipped with tensor cores (TCs) that are commonly used for matrix multiplication in artificial intelligence workloads. However, because they have high computational throughput, they can lead to significant performance gains in other algorithms if they can be successfully exploited. We examine using TCs to compute Euclidean distance calculations, which are used in many data analytics applications. Prior work has only investigated using 64~bit floating point (FP64) data for computation; however, TCs can operate on lower precision floating point data (i.e., 16 bit matrix multiplication and 32 bit accumulation), which we refer to as FP16-32. FP16-32 TC peak throughput is so high that TCs are easily starved of data.  We propose a Fast and Scalable Tensor core Euclidean Distance (\ouralg) algorithm. To achieve high computational throughput, we design  \ouralg for significant hierarchical reuse of data and maximize memory utilization at every level (global memory, shared memory, and registers).  We apply \ouralg to the application of similarity searches, which typically employ an indexing data structure to eliminate superfluous Euclidean distance calculations. We compare to the state-of-the-art (SOTA) TC Euclidean distance algorithm in the literature that employs FP64, as well as to two single precision (FP32) CUDA core algorithms that both employ an index. We find that across four real-world high-dimensional datasets spanning 128-960 dimensions, the mixed-precision brute force approach achieves a speedup over the SOTA algorithms of 2.5--51$\times$. We also quantify the accuracy loss of our mixed precision algorithm to be $<$0.06\% when compared to the FP64 baseline.

\end{abstract}



\keywords{CUDA, Euclidean Distance, GPU, Mixed Precision Floating Point, Self-Join, Similarity Search, Tensor Cores}


\maketitle

\section{Introduction}\label{sec:intro}

Euclidean distance calculations are fundamental to numerous application domains (e.g., in the area of data analytics, they are employed for distance similarity searches~\cite{gowanlock2023optimization}, outlier detection~\cite{zimek2012survey}, k-nearest neighbor searches~\cite{Samet2008}, and clustering~\cite{bottesch2016speeding}). Given the numerous algorithms that rely on Euclidean distance subroutines, improving the performance of this fundamental operation can have a significant impact on a wide range of algorithms and applications.

In this paper, we present a Fast and Scalable Tensor core Euclidean Distance (\ouralg) algorithm, that is designed for moderate to high-dimensional datasets.\footnote{The source code is publicly available: \url{https://github.com/bwcurless/fasted}.} We consider comparing all points in an input dataset to each other, which has a quadratic time complexity, and only returning those points that are within a search distance, $\epsilon$, of each other. This is a common application scenario, as it is assumed in many of the data analytics applications described above. \ouralg can be employed as a subroutine in algorithms and workloads requiring the large scale computation of Euclidean distances.

We define peak throughput as the maximum number of Tera Floating Point Operations Per Second (TFLOPS), which is the maximum throughput achievable based on TC hardware constraints. The Nvidia A100 GPU that we use in our evaluation has a peak FP32 throughput on CUDA cores of 19.5 TFLOPS, whereas TCs can achieve a throughput of 312 TFLOPS using FP16-32~\cite{A100}.\footnote{We use CUDA terminology throughout this paper.} Given that the peak throughput of TCs outlined above is a factor of 16 greater than CUDA cores, it is clear that if they can be efficiently exploited to compute Euclidean distances, there is potential for significant performance improvements over methods that employ CUDA cores.

Based on our investigation, we find that there are two major challenges that limit the performance achievable on GPU TCs for Euclidean distance calculations that we describe as follows.
\begin{enumerate}[wide=0pt]
\item \textbf{Memory Bottlenecks --} A major challenge is efficiently transferring data between each level of the GPU's memory hierarchy, from global memory to shared memory and then into registers. 

\item \textbf{Insufficient Data Reuse --} Given the enormous disparity between FP16-32 TC throughput and global memory bandwidth, exploiting data reuse is critical. 
\end{enumerate}

This paper makes the following contributions:

\begin{enumerate}[wide=0pt]
\item When copying matrix data into streaming multiprocessors (SMs), we employ inline Parallel Thread Execution (PTX) instructions to design efficient memory access patterns.
\item We calculate the data reuse required to maximize performance and design warp- and block-level optimizations to meet these prerequisites. 
\item We design a two-stage pipeline with asynchronous memory copies for loading data from global into shared memory, such that each warp overlaps TC computation with asynchronous loads. 
\item We compare \ouralg to both TC and CUDA core state-of-the-art (SOTA) algorithms and demonstrate that it is superior in all experimental scenarios. 

\end{enumerate} 

The paper is organized as follows: Section~\ref{sec:background} provides relevant background information on similarity searches and an exploration of related work; Section~\ref{sec:our_algorithm} describes our proposed algorithm, \ouralg; Section~\ref{sec:evaluation} presents the experimental evaluation; and, finally, Section~\ref{sec:conclusion} reflects on the work's central contributions and future research directions.

\section{Background}\label{sec:background}

\subsection{Definitions and Problem Statement}\label{sec:problem_statement}
\noindent\textbf{Definition of a dataset and data points --} Let $D$ be a dataset containing $|D|$ points, where each point is defined as $p_i\in D$, and where $i{=}1, 2, \ldots, |D|$. Each point, $p_i$, is defined by coordinates in $d$ dimensions. We define the coordinates for point $p_i$ as $p_{i,k}$, where $k{=}1, 2, \ldots, d$.

\noindent\textbf{Definition of the Euclidean Distance --}  
Let $p_i\in D$ and $p_j \in D$ be two points in the dataset, $D$, where $p_{i,k}$ and $p_{j,k}$ refer to the $k^{th}$ coordinates of points $p_i$ and $p_j$, respectively. The Euclidean distance between $p_i$ and $p_j$ is typically defined as follows and an expansion yields a form that can be computed on TCs~\cite{gallet2022leveraging}:

{
\addtolength{\belowdisplayskip}{-1.0ex}
\addtolength{\belowdisplayshortskip}{-0.5ex}
\addtolength{\abovedisplayskip}{-2.0ex}
\addtolength{\abovedisplayshortskip}{-2.0ex}
\begin{equation}
	\begin{aligned}
		dist(p_i, p_j) &= \sqrt{\sum^{d}_{k=1}(p_{i,k} - p_{j,k})^2} \\
									 &= \sqrt{\sum^{d}_{k=1}p_{i,k}^2 - 2p_{i,k}p_{j,k} + p_{j,k}^2}.
	\end{aligned}
\label{eqn:euclidean_distance}
\end{equation}
}

\noindent\textbf{Definition of queries and the distance similarity self-join --} In this paper, we compare all points $p_i \in D$ to each other, which is a high throughput operation that is well-suited to the GPU. In database terminology, this is referred to as a self-join, where conceptually $D$ can be considered a database table that is joined with itself. This is a typical application scenario for the data analytics applications described in Section~\ref{sec:intro}.

\subsection{Scenarios for Euclidean Distance Calculations}
We describe two common application scenarios for the calculation of Euclidean distances below.

\subsubsection{Scenario 1: Brute Force Distance Similarity Searches}\label{sec:scenario1_brute_with_epsilon}
The distance is calculated between all permutations of points in a dataset $D$ (containing $|D|$ points), and only the pairs where the distance is $\leq\epsilon$ are returned. This has a time complexity of $O(|D|^2)$.

\subsubsection{Scenario 2: Index-Supported Distance Similarity Searches}\label{sec:scenario2_index}
To improve the performance of brute force similarity searches (described above), algorithms will typically first construct an indexing data structure. The index is used to prune the search between points that are sufficiently far from each other, decreasing the amount of comparisons required. 

\subsection{GPU Tensor Cores}\label{sec:background_TCs}

In this section, we discuss the operation of TCs and how they are programmed on Nvidia GPUs. TCs can operate on different floating point data types including FP64, FP32, and FP16-32. There are several libraries that directly employ TCs for linear algebra operations, including cuBLAS and CUTLASS.\footnote{cuBLAS:  \url{https://developer.nvidia.com/cublas}; CUTLASS: \url{https://github.com/NVIDIA/cutlass}} These libraries are suitable when their subroutines can be directly embedded into an algorithm. However, they are unsuitable for our purposes, as they do not offer sufficient granularity or control. There are two other ways to use TCs that provide more flexibility: the WMMA API~\cite{wmmaNvidia}, and PTX instructions.\footnote{\url{https://docs.nvidia.com/cuda/archive/12.0.1/cuda-c-programming-guide/\#element-types-and-matrix-sizes}}

The WMMA API and PTX instructions both load data from shared memory into registers that are shared across an entire warp to compute small submatrices (called register fragments). However, the WMMA API is limited to larger matrix sizes (see Table~\ref{tab:WMMA_vs_PTX}), does not specify the register layout, and yields less control over memory addressing. In this paper, we use PTX instructions as they provide the most flexibility to achieve the highest throughput.

\begin{table}[!t]

\footnotesize
    \centering
		\caption{Comparison of matrix sizes for FP16-32.}
		\begin{tabular}{>{\raggedright\arraybackslash}p{3cm}|p{2.2cm}|p{2.2cm}} 
        \hline
        \rowcolor{blue!40!white}Size ($m$-$n$-$k$)&WMMA API&\ptxmma\\\hline
        16$\times$16$\times$16&\checkmark&\\
        32$\times$8$\times$16&\checkmark&\\
        8$\times$32$\times$16&\checkmark&\\
        8$\times$8$\times$4&&\checkmark\\
        16$\times$8$\times$8&&\checkmark\\
        \textbf{16$\times$8$\times$16} (Used by \ouralg)  &&\checkmark\\
        \hline
        \end{tabular}
    \label{tab:WMMA_vs_PTX}
\end{table}

\begin{lstlisting}[label=lst:ldmatrix, floatplacement=t, basicstyle=\footnotesize, xleftmargin=0.35cm, caption=A 16$\times$16 ldmatrix PTX instruction.]
asm volatile(
"ldmatrix.sync.aligned.x4.m8n8.shared.b16"
"{%0, %1, %2, %3}, [%4];"
: "=r"(A[0]), "=r"(A[1]), "=r"(A[2]), "=r"(A[3])
: "r"(smem_ptr));
\end{lstlisting}

\begin{lstlisting}[label=lst:mma, floatplacement=t, basicstyle=\footnotesize, xleftmargin=0.35cm, caption=A 16$\times$8$\times$16 mma.sync PTX instruction.]
asm volatile(
"mma.sync.aligned.m16n8k16.row.col.f32.f16.f16.f32"
"{%0, %1, %2, %3}, "
"{%4, %5, %6, %7}, "
"{%8, %9}, "
"{%10, %11, %12, %13};"
:"=f"(D[0]), "=f"(D[1]), "=f"(D[2]), "=f"(D[3])
:"r"(A[0]), "r"(A[1]), "r"(A[2]), "r"(A[3]),
"r"(B[0]), "r"(B[1]), 
"f"(C[0]), "f"(C[1]), "f"(C[2]), "f"(C[3]));

\end{lstlisting}

There are two main PTX instructions for using TCs. The first is \texttt{ldmatrix}, shown in Listing~\ref{lst:ldmatrix}, which creates a fragment of a matrix that is stored in registers shared across a warp. The second is \texttt{mma.sync}, shown in Listing~\ref{lst:mma}, which computes $\bm{D}=\bm{A}\times \bm{B}+\bm{C}$, where $\bm{A}$, $\bm{B}$, $\bm{C}$, and $\bm{D}$ are each fragments that were created by prior \texttt{ldmatrix} instructions. 
When using FP16-32 instructions, the data stored in matrices $\bm{A}$ and $\bm{B}$ are represented in FP16 as 16$\times$16 and 16$\times$8 fragments, while the accumulator matrices, $\bm{C}$ and $\bm{D}$, are represented in FP32 as 16$\times$8 fragments.

\subsection{Literature on Expanding TC Applicability}
There are few papers in the literature that extend the applicability of GPU TCs. Most of the papers below are not directly related to this paper, as they examine differing use cases for TCs; however, we summarize them for context.

A pioneering paper by Dakkak et al.~\cite{dakkak2019accelerating} used TCs to perform scan and reduction operations. They show that their reduction algorithm on an Nvidia V100 GPU achieves a 100$\times$ speedup over using CUDA cores.  Ji and Wang~\cite{ji2021accelerating} used TCs to accelerate the DBSCAN clustering algorithm. In their paper, they use cosine distance matrices to determine the $\epsilon$-neighborhood of points when clustering. They achieve a 2.61$\times$ speedup using TCs over CUDA cores. Li et al.~\cite{li2021tcfft} use TCs to improve the performance of half precision Fast Fourier Transforms (FFTs) and achieve a speedup of 3.03$\times$ over the cuFFT library.\footnote{\url{https://developer.nvidia.com/cufft}} Wan et al.~\cite{wan2022novel} used TCs to accelerate the Number Theoretic Transform (NTT) operation that is a bottleneck in the post-quantum cryptography algorithm, CRYSTALS-Kyber~\cite{bos2018crystals}. They demonstrate that their TC NTT algorithm yields a speedup of 6.47$\times$  over the CUDA core algorithm. 

Cui~\cite{cui2024acceleration} proposes using TCs for finite element methods (FEM), which are computationally expensive and thus a good target for GPU TCs. Similarly to Cui~\cite{cui2024acceleration}, we use PTX instructions to eliminate bank conflicts when copying data between registers and shared memory. Unlike  Cui~\cite{cui2024acceleration}, we optimize copying data from global memory into shared memory in a bank-conflict-free manner.

\subsection{Related Work on using Tensor Cores for Euclidean Distance Calculations}\label{sec:background_tedjoin}
To our knowledge, there is one paper in the literature on employing TCs to compute Euclidean distances. The paper proposes ``Tensor Euclidean Distance Join'' (\tedjoin)~\cite{gallet2022leveraging}, which is used to compute the Euclidean distances between all points in a dataset for similarity searches (the two scenarios outlined in Sections~\ref{sec:scenario1_brute_with_epsilon}--\ref{sec:scenario2_index}). 

\tedjoin~\cite{gallet2022leveraging} is designed for FP64 calculations and uses the WMMA API. It was found to perform best when the data dimensionality ($d$) is low rather than high. Given that \tedjoin is the only other TC algorithm in the literature, we compare \ouralg to \tedjoin in our evaluation.

\subsection{Literature on Distance Similarity Searches}\label{sec:related_work_gpu_similarity_searches}
In this section, we discuss the literature on the distance similarity self-join as defined in Section~\ref{sec:scenario2_index}. \gdsjoin~\cite{gowanlock2019gpu,gowanlock2023optimization} is a CUDA core GPU algorithm that uses a grid-based index for the distance similarity self-join.  
The algorithm uses several optimizations, including assigning points to threads within warps such that they have low intra-warp load imbalance, processing warps from largest to smallest workload to prevent inter-warp load imbalance towards the end of a kernel execution, and reordering the coordinates in the dataset to increase the probability that when computing the distance between points, the distance calculation will abort early (we refer to this as short circuiting).

\mistic~\cite{donnelly2024multi} is another CUDA core distance similarity search algorithm that is similar to \gdsjoin. A major difference between \gdsjoin and \mistic is that the latter uses a combination of metric- and coordinate-based partitioning to prune the search for nearby points of each query point. Additionally, \mistic uses incremental index construction to select between candidate partitions. \mistic outperforms \gdsjoin across all experiments (on datasets up to $d{=}90$). Interestingly, the authors find that one reason \mistic outperforms \gdsjoin is that it has better load balancing properties across warps and blocks executing on SMs.  The algorithm we propose in this paper, \ouralg, has perfect load balancing, which indicates that it could have a performance advantage over these CUDA core algorithms.

In Section~\ref{sec:evaluation}, we compare \ouralg to \gdsjoin and \mistic, as they are SOTA CUDA core algorithms. \ouralg does not use an index, so it is disadvantaged compared to these approaches. We do not compare to CPU algorithms in this paper, as the GPU algorithms are significantly faster~\cite{gowanlock2019gpu,gowanlock2023optimization,gallet2022leveraging}, particularly when processing high-dimensional datasets.

\section{\ouralg}\label{sec:our_algorithm}
We present our Fast and Scalable Tensor core Euclidean Distance (\ouralg) algorithm, which exploits (FP16-32) TCs on GPUs to compute Euclidean distances. We begin by describing how TCs can be used for Euclidean distance calculations and then describe the optimizations required to effectively utilize FP16-32 TCs using the A100 as our platform (the method is generalizable to other TC-equipped GPU models as well).

\subsection{Overview of Euclidean Distance Calculations on Tensor Cores} \label{sec:eucliddisttc_explanation}

Figure~\ref{fig:single_mma} shows an illustrative example of calculating Euclidean distances, where the distances between points $p_1, \ldots,p_{16}$ and $p_1,\ldots,p_8$ are computed.  The steps below describe how TCs can be used to calculate the resulting set of 128 Euclidean distances:

\noindent
{\bf Step~1:} On CUDA cores, we precompute the sum of the squared dimensions, $s_i$, for each point $p_i \in D$, $s_i = \sum\limits_k{p_{i,k}^2}$ and store them in global memory. All summations round towards zero to match TC rounding~\cite{tcRounding}.

\noindent
{\bf Step~2:} Tensor cores operate on small matrices that are stored in registers; we refer to these matrices as ``register fragments,'' or ``rf''. There are three register fragments required to compute distances:
\begin{enumerate}[wide=0pt]
\item\prf: 16 dimensions from 16 points in dataset $D$. We refer to these simply as the ``points''. 
\item\qrf: 16 dimensions from 8 points in dataset $D$ in a transposed layout. We refer to these as the ``query points''. 
\item\arf: A $16\times 8$ fragment initialized with zeros.
 \end{enumerate}
 We use TCs to multiply and accumulate these three fragments to obtain: $\arf = \prf \times \qrf + \arf$. As TCs can only compute inner products of $16$ dimensions per operation (see Table~\ref{tab:WMMA_vs_PTX}), we must iteratively load new point data into \prf\ and \qrf\ and accumulate into \arf\ $\frac{d}{16}$ times. 

\noindent
{\bf Step~3:} On CUDA cores, we combine the elements of \arf\ ($a_{i,j}$), along with $s_i$ and $s_j$ (step~1) to compute the square of the final distance (from Equation~\ref{eqn:euclidean_distance}) $dist_{i,j}^2 = -2 \times a_{i,j} + s_i + s_j$.

\subsection{Overview of the Data Path from Global Memory to Registers}\label{sec:mptc_overview_data_path}

While computing Euclidean distances using TCs is conceptually straightforward, computing step~2 in Section~\ref{sec:eucliddisttc_explanation} with high throughput is challenging. Previous GPU work that computed distance similarity searches focused on algorithmic optimizations using indexed search methods to reduce the search space, as well as short circuiting techniques to reduce computation (Section~\ref{sec:background_tedjoin}--\ref{sec:related_work_gpu_similarity_searches}). In contrast, maximizing performance on TCs requires a balanced workload that these index-supported algorithms cannot provide (e.g., they have non-deterministic execution pathways that cause warp divergence). We propose several optimizations to balance the workload and maximize TC throughput.

\begin{figure}
    \centering
    \includegraphics{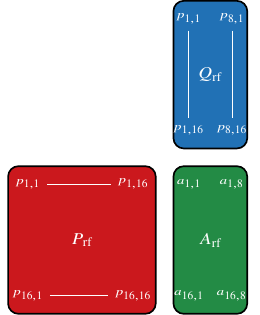}
    \caption{A single 16$\times$8$\times$16 TC computation used to accumulate the distances between two points $p_i$ and $p_j$ into $a_{i,j}$. Only the corner elements in each matrix are shown for illustrative purposes.}
    \label{fig:single_mma}
		\Description{Diagram showing how a 16 by 16 P fragment multiplied by a 16 by 8 Q fragment yields a single 16 by 8 A fragment.}
\end{figure}

\begin{figure}
    \centering
    \includegraphics{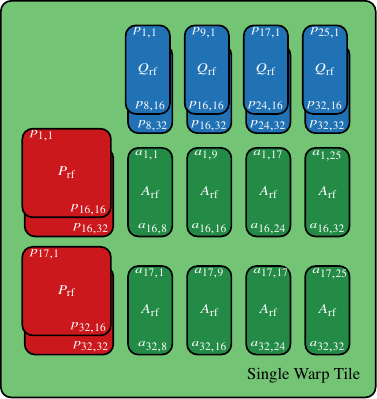}
    \caption{Two iterations of a $32{\times} 32{\times }16$ warp tile are shown. It is made up of many register fragments from Figure~\ref{fig:single_mma}. Only the upper left and lower right coordinates of each fragment are specified for illustrative purposes.}
    \label{fig:warp_mma}
		\Description{Diagram showing how single MMA register fragments can be combined to compute an entire warp tile.}
\end{figure}

\begin{figure}
    \centering
    \includegraphics{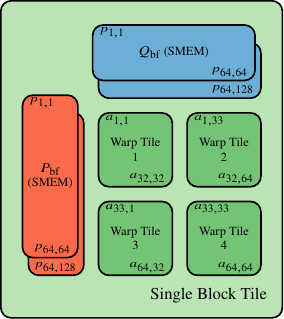}
    \caption{Two iterations of a $64{\times }64{\times }64$ block tile are shown. Two \sixfourd block fragments, \pbf \ and \qbf, are paged from global memory to shared memory (SMEM). Four warp tiles (Figure~\ref{fig:warp_mma}) each page a portion of the block fragments into their own register fragments and accumulate the distances. As $d{=}128$, this sequence is repeated for two iterations.}
    
    \label{fig:block_tile_mma}
		\Description{Diagram showing how four warp tiles can all share data in shared memory to compute a single block tile.}
\end{figure}

Computing a large matrix of Euclidean distances with \ouralg is shown in Figures~\ref{fig:single_mma}-\ref{fig:entire_mma} and we describe them as follows. 

The foundation is a single PTX instruction (shown in Listing~\ref{lst:mma} and Figure~\ref{fig:single_mma}) that computes a $16{\times} 8{\times} 16$ ``$k$-slice'' of distances. A \kslice refers to a contiguous subset of a point's total dimensions (e.g. four \sixtd \kslices together represent a \sixfourd point).

Transferring data from global memory into registers for the MMA PTX instruction is a bottleneck. Box~\#1 estimates the amount of data that needs to be reused when reading from global memory and reading from shared memory in order to reach peak throughput. As will be described below, this estimation derives tile size prerequisites that ensure conditions (1) \& (2) from Box~\#1 are met.

\begin{titled-frame}{Box \#1: Selecting Tile Sizes Based on Data Reuse}

In a matrix multiply-accumulate product, there are two floating-point operations for every two elements processed. Peak throughput (312~TFLOPS) requires reading the following number of elements per second:
\[
\frac{\text{312 TFLOP}}{\text{second}} \times\frac{\text{2 elements}}{\text{2 FLOP}} = \frac{\text{312}\times 10^{12}~\text{elements}}{\text{second}}.
\]

{\bf (1)} Global memory bandwidth is 1.5 TB/s. With a 100\% hit rate, the L2 cache increases this to 6.4~TB/s. Therefore each value read from global memory must be reused:
\[
\frac{\text{312}\times 10^{12}~\text{elements}}{\text{second}} \times \frac{\text{2~B}}{\text{1 element}} \times \frac{\text{second}}{6.4~\text{TB}} = \textbf{98 times}.
\]

{\bf (2)} Shared memory bandwidth is 17.9 TB/s. Therefore each value read from shared memory must be reused:
\[
\frac{\text{312}\times 10^{12}~\text{elements}}{\text{second}} \times \frac{\text{2 B}}{\text{1 element}} \times \frac{\text{second}}{\text{17.9 TB}} = \textbf{35 times}.
\]
\end{titled-frame}

We create a ``warp tile'' (Figure~\ref{fig:warp_mma}) that contains multiple \prf\ and \qrf\ fragments. Each \prf\ is multiplied by every \qrf, and the results are accumulated into the corresponding \arf . Only one \sixtd \kslice can fit into registers at a time, so fragments are loaded and accumulated into \arf\ in four separate iterations. This sequence computes a $64{\times}64{\times}64$ warp tile.

There are four schedulers and four TCs per SM, so we elect to run four warp tiles simultaneously on each SM. These four warps form a ``block tile'' (Figure~\ref{fig:block_tile_mma}) that computes a $128{\times}128$ tile of distances $dist(p_i,p_j)^2$. A block tile transfers two different ``block fragments,'' which are \sixfourd \kslices of $128$ points (\pbf \ and \qbf), from global memory to shared memory that each warp tile reads from during its four iterations. This repeats until all dimensions of the points have been accumulated by the warp tiles and the final distances have been computed. When a block completes a tile, a work queue provides the next tile to compute and this process repeats (Figure~\ref{fig:entire_mma}).

\subsection{Optimizations: Mitigating the Memory Bottleneck}\label{sec:ouralg_optimizations}

\subsubsection{L2 Cache: Maximizing Data Reuse by Ordering Block Tiles}
\label{opt:l2cache}

\begin{figure}
    \centering
    \includegraphics{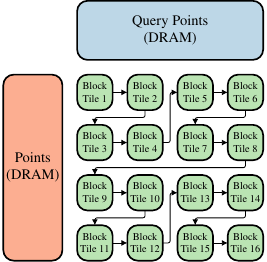}
    \caption{An illustration of how a work queue orders block tile computations into small squares to improve the L2 cache hit rate when reading from global memory (DRAM). }
    \label{fig:entire_mma}
		\Description{Diagram showing how many block tiles can share the same data from global memory by computing subcomponents of the larger matrix in small square tiles.}
\end{figure}

The L2 cache has a bandwidth of 6.4 TB/s which is $\sim$4$\times$ greater than global memory. Our algorithm is able to utilize most of this bandwidth by having blocks retrieve work from a queue such that concurrently executing block tiles read similar data elements (point coordinates) to maximize spatial locality across SMs (Figure \ref{fig:entire_mma}). 

\subsubsection{Single Block Tile Shared Memory Buffering}\label{opt:blocktilesmembuffering}

 Every block tile is responsible for computing a 128$\times$128 tile of the distance matrix. A \sixfourd \kslice of the input data is stored in shared memory and used by all four warps. This block tile size yields $128\times$ data reuse which is more than that needed to reach peak TC throughput (Box~\#1).

\subsubsection{L1 Cache elimination}\label{sec:L1_elimination}
Each SM has 192~KB of memory that can be used as an L1 cache or directly manipulated as shared memory. In order to maximize throughput, most of the L1 cache has been configured as shared memory to allow us to use a pipelined execution that hides memory access latency, which will be described in the following sections.

\subsubsection{Asynchronous Global to Shared Memory Transfers}\label{opt:async_global_shared_memory}
Normal synchronous methods of transferring data from global to shared memory use synchronous copies where the data path is as follows: global memory $\rightarrow$ L2 Cache $\rightarrow$ L1 Cache $\rightarrow$ registers $\rightarrow$ shared memory. The A100 GPU supports asynchronous memory copies~\cite{A100} using \texttt{cuda::memcpy\_async} that skips the L1 cache and registers, storing directly into shared memory. This reduces latency and register pressure.

\subsubsection{Global to Shared Memory Transfer Pipelining}
\label{opt:pipelining}
There is sufficient shared memory to create a multi-stage pipeline of memory transfers using \texttt{cuda::pipeline} with \texttt{cuda::memcpy\_async} as described in Section~\ref{opt:async_global_shared_memory}. Our algorithm issues asynchronous memory copies in a two-stage pipeline and overlaps computation with data transfer to hide latency.

\subsubsection{Multiple Blocks Per Streaming Multiprocessor}\label{opt:multipleblockspersm}
We use multiple blocks that are executed in parallel on each SM to hide high latency instructions like MMA and memory transfers. We elect to maximize the following: block tile size, \kslice width, and pipeline depth (two levels) while leaving sufficient shared memory and registers to allow two blocks to run simultaneously on each SM.

\subsubsection{Single Warp Tile}
\label{opt:warptile}
We create a $64{\times }64{\times }16$ warp tile (Figure \ref{fig:warp_mma}) that reuses each \prf\ and \qrf\ fragment, $4$ and $8$ times respectively, to achieve the required reuse calculated in Box~\#1. To reduce register pressure, only a single \sixtd \kslice is stored in registers at a time.

\subsubsection{Memory Address XOR Swizzling}\label{opt:swizzling}
\begin{figure}[!t]
    \centering
    \includegraphics{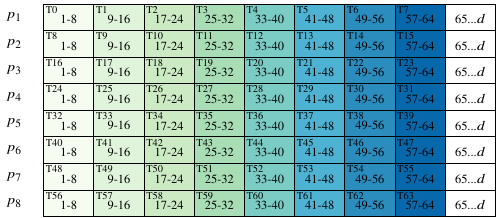}
    \caption{The row-major global memory layout for the first 8 points is shown here. The threads that are responsible for reading each 8 dimension chunk of data from global memory and storing it in shared memory are labeled in the top left corner of each rectangle (T0--T63).}
    \label{fig:global_memory}
		\Description{Global memory layout showing how data is paged from global memory in row-major format, in groups of 64 dimensions.}
\end{figure}

\begin{figure}[!t]
    \centering
    \includegraphics{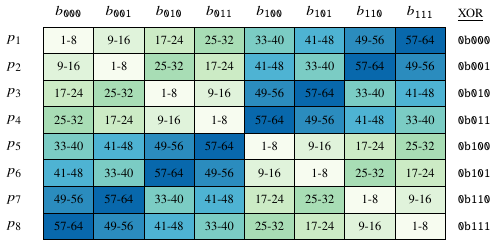}
    \caption{The point data from Figure~\ref{fig:global_memory} is copied into the shared memory layout shown here one row at a time. The destination bank, $b$, of each 8 dimension chunk is swizzled by computing the XOR of the original bank $b$ with the row number (shown on the right). Each column in the table represents a group of 4 of the 32 shared memory banks. Reordering the data makes the \texttt{ldmatrix} instruction, shown in Figure~\ref{fig:ldmatrix}, access unique banks in each phase of its operation. }
    \label{fig:shared_memory}
		\Description{Shared memory layout showing how chunks of 8 dimensions from the global memory layout are swizzled into different banks of shared memory to create conflict free accesses.}
\end{figure}

\begin{figure}[!t]
    \centering
    \subfigure[]{
      \includegraphics{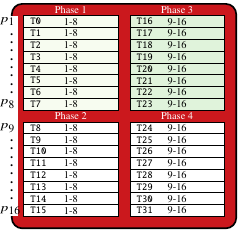}
    }
    \subfigure[]{
      \includegraphics{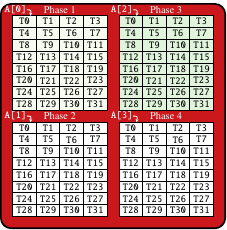}
    }
    \caption{(a) Single fragment \prf\ shared memory access layout. (b) Single fragment \prf\ register layout. The result of a single $16{\times}16$  \texttt{ldmatrix} instruction reading from shared memory and storing into a fragment made up of registers is shown above. Each 8 dimension chunk is labeled with the thread responsible for reading it from shared memory. After each thread reads its 16~B chunk from shared memory, that data is shuffled into the registers of four different threads (b). For example, T0 reads $p_1$ dimensions $k{=}1{-}8$, but it is stored into four registers, owned by threads 0-3. }
    \label{fig:ldmatrix}
		\Description{Register layout showing how the 8 dimension chunks of data are ready from shared memory, and stored into registers that all threads in the warp share.}
\end{figure}

When paging data into and out of shared memory, we ensure  that all global reads are fully coalesced, while simultaneously eliminating shared memory bank conflicts.  To achieve this, the shared memory addresses are swizzled as values are stored (using XOR), and unswizzled as they are read into registers. Figures~\ref{fig:global_memory}--\ref{fig:ldmatrix} illustrate the global memory, shared memory, and register layouts, respectively, showing the path that data takes when being read from global memory and stored into the registers corresponding to a single fragment.

The FP16 point data is stored in global memory in row-major order with each point having 128 B alignment. The first \sixfourd of the first 8 points are shown in Figure~\ref{fig:global_memory}. Groups of 8 threads work together to load a single \sixfourd \kslice of a point from global memory and store it in shared memory.

Shared memory contains 32 discrete 4~B banks. Figure~\ref{fig:shared_memory} shows which shared memory bank each \eightd slice of point data was stored in when loading from global memory (Figure~\ref{fig:global_memory}). The destination address is ``swizzled,'' meaning that for a given \eightd slice $s$ of a point $p_i$, the destination shared memory address is: 

{
\addtolength{\belowdisplayskip}{-0.5ex}
\addtolength{\belowdisplayshortskip}{-0.5ex}
\addtolength{\abovedisplayskip}{-2.0ex}
\addtolength{\abovedisplayshortskip}{-2.0ex}
\begin{equation}
A_\text{dest} = 8\cdot(i-1) + s \oplus ((i-1)\bmod 8).
\end{equation}
}

Figure~\ref{fig:shared_memory} shows how each \eightd slice in a group of 8 threads will be stored in a unique bank and is thus free of bank conflicts.

While swizzling the data layout is not required to have coalesced and bank-conflict-free shared memory stores, it is imperative for loading the point data from shared memory and storing it into registers. To load a single fragment \prf, there are four phases, or memory transactions, in a \texttt{ldmatrix} instruction (see Listing~\ref{lst:ldmatrix} and Figure~\ref{fig:ldmatrix}). Each phase generates a single 128~B memory transaction that must be bank-conflict-free to achieve maximum throughput. 

To see that banks are accessed in a conflict-free manner, observe that each \eightd slice of data being accessed in a single phase in Figure~\ref{fig:ldmatrix} is located in a unique bank in Figure~\ref{fig:shared_memory}. In contrast, a simple row-major copy from global memory (Figure~\ref{fig:global_memory}) into shared memory would create 8-way bank conflicts during each phase.

\subsubsection{Shared Memory Alignment}\label{opt:smemalignment}
Shared memory allocations do not have alignment guarantees. 
Alignment specifiers, \texttt{\_\_align\_\_\-(128)}, were added to shared memory definitions to ensure 128~B alignment. This enables swizzling to place the data in the appropriate bank, thus mitigating bank conflicts.

\section{Experimental Evaluation}~\label{sec:evaluation}

\subsection{Experimental Methodology}\label{sec:exp_method}
\subsubsection{Platform}
We conduct our experiments on a platform with 2$\times$ AMD EPYC~7542 CPUs (64 physical cores total) clocked at 2.9 GHz with 512 GiB of main memory containing an Nvidia A100 GPU with 40 GiB of global memory (PCIe model). The host code uses C/C++ and the \ouralg GPU code is parallelized using CUDA v.12.6.3. All host code is compiled with the O3 optimization flag. Reported response times are averaged over three trials. In all experiments using \ouralg or the reference implementations, the variation between time trials is negligible, such that the error bars are smaller than the data points on the plots, so we omit plotting these statistics. However, for completeness, we include this information in the associated figures for \ouralg.

We examine two main metrics when measuring performance. \emph{Response time:} We measure the response time of the algorithms. \emph{Derived TFLOPS:} We report the TFLOPS yielded by the TCs by taking the total number of operations and dividing it by the measured response time. We refer to this when we discuss TFLOPS.

To make a fair comparison between all implementations, we include all overheads, such as allocating memory, transferring data to and from the GPU, and storing the results in main memory. Regarding methods that employ indexing, we include all index construction overheads.

\subsubsection{Implementations}\label{sec:implementation_configurations}
We compare \ouralg to several GPU implementations. In what follows, we outline the reference implementations, their configurations, and the default configuration for \ouralg. We summarize the implementations in Table~\ref{tab:implementations}.

\noindent\textbf{\ouralg \ --} We configure \ouralg using the parameters outlined in Table~\ref{tab:optimized_parameters}, which were motivated in Section~\ref{sec:our_algorithm}.

\begin{table}[!t]
\footnotesize
\caption{Summary of optimized parameters. }
\begin{tabular}{p{5.4cm}|r} 
        \hline
        \rowcolor{blue!40!white}Parameter & Optimized Value \\ \hline
        Block tile dispatch shape & $8{\times }8$ blocks \\ \hline
        Block tile iteration size & $128{\times }128{\times }64$ \\ \hline
        Number of blocks in grid& $2{\times}$ \# of SMs (216 total) \\ \hline
        Warp tile iteration size & $64{\times }64{\times }16$ \\ \hline
        Warps per block & 4 \\ \hline
        Pipeline depth & 2 \\ \hline
        \end{tabular}
    \label{tab:optimized_parameters}
\end{table}

\noindent\textbf{\tedjoin \ --} As described in Section~\ref{sec:background_tedjoin}, \tedjoin~\cite{gallet2022leveraging} is the only other algorithm in the literature that computes Euclidean distances on TCs, and it uses double precision floating point (FP64) values.

\tedjoin exceeds the shared memory capacity and fails to compile when a dataset contains points with $d>128$. We elected to modify \tedjoin to allocate part of the L1 cache as shared memory to allow us to employ datasets with $d\leq 384$. \tedjoin can operate in two modes, the first using brute force searches, and the second using an index for index-supported range queries. We refer to these variants as {\bf \tedjoinbrute} and {\bf \tedjoinindex}, respectively.

\noindent\textbf{\gdsjoin \ --} As described in Section~\ref{sec:related_work_gpu_similarity_searches} this algorithm performs index-supported range queries using CUDA cores. We use  the same optimizations and configuration as described in the \gdsjoin papers~\cite{gowanlock2019gpu,gowanlock2023optimization}, so we only outline the differences here. We configure \gdsjoin to use FP32 data, which provides a baseline of comparison to \ouralg that does not use FP64 like \tedjoin. Lastly, we use a batch size of $b_s=2\times10^9$ for batching the result set from the GPU to the host. We also employ \gdsjoin as executed using FP64 to compute the accuracy of \ouralg's mixed-precision results. The code is publicly available.\footnote{\url{https://github.com/mgowanlock/gpu_self_join}} 

\noindent\textbf{\mistic \ --} This algorithm is similar to \gdsjoin above as it uses CUDA cores (Section~\ref{sec:related_work_gpu_similarity_searches}). We use the same configurations and optimizations as outlined in the \mistic paper~\cite{donnelly2024multi} and use a block size of 256 with 1024 blocks per kernel invocation (multiple invocations are used to batch the result set from device to host). The algorithm uses 6 levels and selects between 38 candidate layers at each level when incrementally constructing the index. \mistic is configured to use FP32 instead of FP64 (the latter was used in the evaluation in the \mistic paper~\cite{donnelly2024multi}). The code is publicly available.\footnote{\url{https://github.com/bwd29/self-join-MiSTIC}}

\noindent\textbf{Short circuiting of distance calculations -- }The CUDA core algorithms (\gdsjoin and \mistic) abort computing the distance between each pair of points when the running total distance exceeds $\epsilon$. When points in a dataset are very spread out, this saves significant computation time. The TC algorithms (\tedjoin and \ouralg) are less flexible because they compute the distances between all permutations of two sets of points simultaneously. \ouralg compares an entire 128$\times$128 block of points at a time. As \ouralg is computing 16,384 distances at a time, and every single point would need to short circuit for the optimization to be viable, we forego this optimization. \tedjoin compares an 8$\times$8 matrix of points at a time in FP64 mode, which is small enough that it does employ short circuiting. \tedjoinbrute does not abort early.

\begin{table}[!t]
\caption{Comparison of implementation properties.}
\footnotesize
    \centering
    {
		\begin{tabular}{l|l|l|>{\raggedright\arraybackslash}p{1.2cm}|>{\raggedright\arraybackslash}p{1.8cm}} 
        \hline
        \rowcolor{blue!40!white}Implementation&GPU Core&Precision& Scenario~1: Brute Force  (Sec.~\ref{sec:scenario1_brute_with_epsilon})&Scenario~2: Index-Supported (Sec.~\ref{sec:scenario2_index})\\ \hline
        \ouralg&Tensor&FP16-32&\checkmark&\\
        \tedjoinbrute&Tensor&FP64&\checkmark&\\
        \tedjoinindex&Tensor&FP64&&\checkmark\\
        \gdsjoin&CUDA&FP32&&\checkmark\\
        \mistic & CUDA & FP32 & &\checkmark\\
        \hline
        \end{tabular}
    }
    \label{tab:implementations}
\end{table}

\subsubsection{Datasets \& Selectivity Values}

\noindent\textbf{Real-World Datasets --} We use real-world datasets to compare the performance of \ouralg to reference implementations that employ indexing methods because their performance is impacted by the data distribution and the search radius, $\epsilon$. These datasets are standard benchmarks for evaluating search algorithms on high-dimensional datasets~\cite{xie2018image,zheng2017sift,lu2017deep,ryali2020bio,huang2023lightweight}, are publicly available\footnote{\url{https://www.cse.cuhk.edu.hk/systems/hash/gqr/datasets.html}}, and are summarized in Table~\ref{tab:datasets}.

\noindent\textbf{Synthetic Datasets --} Table~\ref{tab:datasets} outlines the dataset size and dimensionality of the synthetic dataset that we use in the evaluation. We use this class of datasets to compute the throughput of the brute force approaches, which require comparing all points to each other; therefore, the data distribution does not impact performance.

\noindent\textbf{Selectivity of the Range Queries --} The search radius, $\epsilon$, used in a similarity search will impact the degree of pruning afforded by an indexing data structure. A large search radius that returns a large number of neighbors per point in a dataset will reduce the degree of index pruning, and a small value of $\epsilon$ will allow many points to be ignored when performing distance calculations. To standardize experiments across datasets, we employ the selectivity, which refers to the mean number of points found by each point searched in the dataset. The selectivity is defined as $S{=}(|R|{-}|D|)/|D|$, where $|R|$ is the total result set size (the number of pairs of points found within $\epsilon$ of each other). Common selectivity values are used to ensure that searches have a proportional amount of work to each other; otherwise, comparing performance between datasets is less meaningful. In our evaluation, we select values of $\epsilon$ for each dataset that yield three levels of selectivity (small, medium, and large), and these are defined as $S_s{=}64$, $S_m{=}128$, and $S_l{=}256$. The $\epsilon$ values corresponding to these selectivity values are given for the real-world datasets in Table~\ref{tab:datasets}.

\begin{table}[!t]
\caption{\emph{Top:} Real-world datasets that we use in our evaluation. The dataset size, $|D|$, and data dimensionality, $d$, are shown, in addition to the $\epsilon$ values that correspond to three target selectivity values that we employ ($S_{s}$,  $S_{m}$, and $S_{l}$). \emph{Bottom:} Synthetic datasets that we use in our evaluation.}
\footnotesize
    \centering
        \begin{tabular}{l|r|r|r|r|r} 
        \hline
             
            \rowcolor{blue!40!white} \multicolumn{6}{c}{Real-World Datasets}\\\hline
           \rowcolor{blue!40!white}  Dataset& $|D|$ & $d$  & $\epsilon (S_{s})$  &$\epsilon (S_{m})$  &$\epsilon (S_{l})$ \\\hline
            \sift&  10,000,000 &  128 & 122.5 & 136.5 & 152.5 \\ \hline
             \tinyds&  5,000,000 &  384 & 0.1831 & 0.2045 & 0.2275 \\ \hline
             \cifar&  60,000 &  512 & 0.6289 & 0.6591 & 0.6914 \\ \hline
             \gist&  1,000,000 &  960 & 0.4736 & 0.5292 & 0.5937 \\ \hline
           \rowcolor{blue!40!white}  \multicolumn{6}{c}{Synthetic Datasets}\\\hline
          \rowcolor{blue!40!white}   Dataset& \multicolumn{2}{c|}{ $|D|$ }& \multicolumn{3}{c}{$d$}\\\hline
             \synth&\multicolumn{2}{c|}{$10^{3 + n/3} \text{ where } n\in [0, 1, \ldots, 9]$}&\multicolumn{3}{c}{$d\in 2^n \text{ where } n\in[6, 7, \ldots, 12]$}\\\hline
        \end{tabular}
    \label{tab:datasets}
\end{table}

\subsection{\ouralg: Performance as a Function of Dataset Size and Dimensionality}\label{sec:eval_dataset_size_dimensionality}
With all of our optimizations described in Section~\ref{sec:ouralg_optimizations} enabled, we examine how the performance of \ouralg varies as a function of dataset size ($|D|$) and dimensionality ($d$). Some of the optimizations are sensitive to these parameters so we vary them by using the \synth dataset and plot the throughput as derived TFLOPS. We only include the kernel execution times to derive TFLOPS as we are interested in understanding FP16-32 performance without the additional overheads included in the end-to-end response time (e.g., data transfers between host and GPU).

\begin{figure}
    \centering
    \includegraphics[scale=1.0]{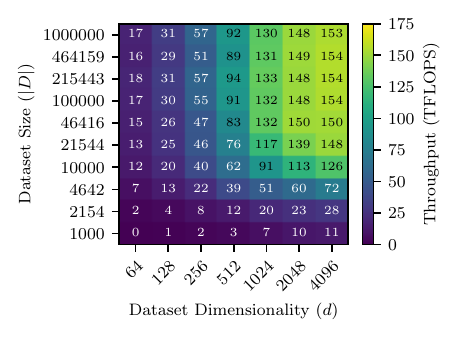}
    \caption{The number of TFLOPS achieved using \ouralg as a function of dataset size ($|D|$) and dimensionality ($d$) on the \synth class of datasets. The maximum throughput was 154 TFLOPS with a standard deviation of 0.02.}
    \label{fig:tflops}
		\Description{Plot showing how dataset dimensionality and size affect the computational throughput. Increasing dimensionality and size increase throughput.}
\end{figure}

Figure~\ref{fig:tflops} shows the results of this experiment. Our maximum throughput is roughly 150 TFLOPS, which can be achieved with a minimum dataset size of $|D|{=}46416$ and dimensionality of $d{=}2048$. In practice, this indicates that only a small to moderate dataset size is needed to reach peak performance, assuming that the dataset contains a sufficient number of dimensions.

Recall that because \ouralg is brute force, its performance is not impacted by the data distribution, so these results apply to any real-world dataset of the same size and dimensionality. To estimate \ouralg's performance on a dataset with a dimensionality that is not a multiple of $64$, one must use the throughput measurements from the next multiple of $64$. In other words, for low dimensionality like $d{=}65$, $63$ dimensions of zeros are added, and the throughput is almost halved. This effect on performance decreases as dimensionality increases.

\subsection{\ouralg: Impact of Optimizations}

\def\baselineTFlops{154}
\def\ltwocache{133.1}
\def\blocktilesmembuffering{95.8}
\def\asyncglobalsharedmemory{48.6} 
\def\pipelining{145.0}
\def\multipleblockspersm{110.8}
\def\warptile{38.0}
\def\swizzling{120.8}
\def\smemalignment{120.7}

In Section~\ref{sec:ouralg_optimizations}, we described numerous optimizations and implementation details that were primarily used to mitigate the memory bottleneck when using FP16-32 TCs. Here, we present the relative impact of those optimizations by performing a leave-one-out evaluation that disables an optimization while leaving the remainder of the optimizations enabled. All optimizations were individually disabled, except for the \texttt{cuda::memcpy\_async} optimization which required disabling the multi-stage pipeline as well.\footnote{ Synchronous memory copies cannot be pipelined using the libcudacxx API: \url{https://nvidia.github.io/cccl/libcudacxx/extended_api/synchronization_primitives/pipeline.html}}  We select the \synth dataset with $|D|{=}10^5$ and $d{=}4096$ because as we showed in Section~\ref{sec:eval_dataset_size_dimensionality}, this dataset size and dimensionality are sufficient to reach our maximum throughput with \ouralg.

From Table~\ref{tab:optimization_comparison}, we find that there are several optimizations that have an exceptional impact on performance, and we outline them as follows: $(i)$ We find that warp tile optimization has the greatest impact because reading data from shared memory into registers and reusing the \kslice data numerous times is needed to ensure that the TCs are not underutilized. $(ii)$ The asynchronous memory copies from global to shared memory are critical to reducing memory latency. $(iii)$ Using a block tile with four warps reusing data from shared memory significantly improves performance.

Despite these three optimizations having an outsized performance impact, none of the optimizations were insignificant, as they were all needed to reach our maximum throughput of \baselineTFlops~TFLOPS.  As we will discuss later, the maximum throughput reported here is a lower bound on the maximum due to the throttling of clock speeds on our GPU.

\begin{table}[!t]
\caption{Performance sensitivity study using the leave-one-out method. We enable all optimizations but leave each of the optimizations out in isolation to determine their relative impact on performance. Experiments are conducted using the \synth dataset with $|D|=10^5$ and $d{=}4096$.}

\footnotesize
    \centering
    {
			\begin{tabular}{p{4.7cm}|l|r} 
        \hline
        \rowcolor{blue!40!white}Disabled Optimization & Section &Derived TFLOPS \\ \hline
        Block Tile Ordering &\ref{opt:l2cache} & \ltwocache \\
        Block Tile &\ref{opt:blocktilesmembuffering} & \blocktilesmembuffering \\
        Memcpy Async \& Multi-stage Pipeline &\ref{opt:async_global_shared_memory}--\ref{opt:pipelining} & \asyncglobalsharedmemory \\
        Multi-stage Pipeline &\ref{opt:pipelining} &\pipelining \\
        SM Block Residency &\ref{opt:multipleblockspersm} &\multipleblockspersm \\
        Warp Tile &\ref{opt:warptile} & \warptile \\
        Swizzled SMEM Layout &\ref{opt:swizzling} &\swizzling \\
        Shared Memory Alignment &\ref{opt:smemalignment} &\smemalignment \\\hline
        All Optimizations Enabled &\ref{sec:ouralg_optimizations}&\baselineTFlops\\
        \hline
        \end{tabular}
    }
    \label{tab:optimization_comparison}
\end{table}

\subsection{Brute Force Tensor Core Throughput}\label{sec:eval_throughput}

\begin{figure}[!t]
    \centering
    \includegraphics[scale=0.95]{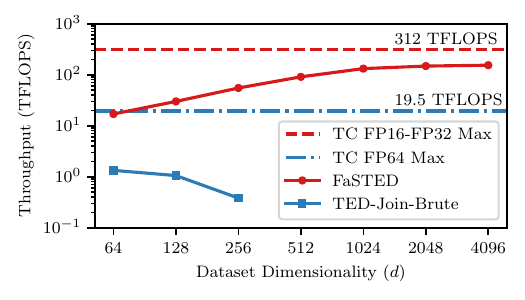}
    \caption{Comparison of derived TFLOPS (plotted on a log scale) achieved on the brute force TC algorithms, \tedjoinbrute (FP64) and \ouralg (FP16-32), as a  function of data dimensionality, using the \synth dataset of size $|D|{=}10^5$.  The standard deviation for all \ouralg measurements was $<0.7$ TFLOPS. The peak throughput is shown for context. }
    \label{fig:tflops_synthetic_comparison}
		\Description{Plot comparing the performance of TED Join and \ouralg relative to the theoretical maximum tensor core throughputs for FP64 and FP16-32. } 
\end{figure}

We compare the performance of the two brute force TC algorithms to observe how they scale with increasing data dimensionality. Recall from Section~\ref{sec:implementation_configurations} that \tedjoinbrute exceeds shared memory capacity and so we are unable to report its performance across a large range of data dimensionalities.  Figure~\ref{fig:tflops_synthetic_comparison} compares the performance of our algorithm, \ouralg (FP16-32), to \tedjoinbrute (FP64). Regarding \ouralg, we find that the TC throughput increases as a function of $d$, which indicates that it is highly scalable in terms of data dimensionality. In contrast, the performance of \tedjoinbrute degrades significantly as $d$ increases, which was also shown in the \tedjoin paper~\cite{gallet2022leveraging}. At $d{=}4096$, \ouralg achieves a sustained throughput of $49\%$ of the A100's peak throughput of 312 TFLOPS using FP16-32 TCs~\cite{A100}. In contrast, \tedjoin only achieves 6.8\% of FP64 peak throughput when $d{=}64$, and it declines at higher dimensionalities.

To better understand why the performance of \ouralg scales well with increasing data dimensionality ($d$), while the performance of \tedjoinbrute degrades, we employ the Nvidia Nsight Compute profiler and present the results in Table~\ref{tab:profiler_results}. We observe that as $d$ increases, the tensor pipe utilization metric increases for \ouralg but does not for \tedjoinbrute. Examining the percentage of shared memory bank conflicts, we observe that \ouralg is conflict-free (0\%), whereas \tedjoinbrute yields $\geq$75\% shared memory bank conflicts. \tedjoin is constrained by the WMMA API, which has rigid load/store memory access patterns, which causes shared memory bank conflicts in both the \tedjoinbrute and \tedjoinindex variants.

While analyzing our results from Table~\ref{tab:profiler_results}, we discovered a discrepancy between the TC Pipe Utilization metric and our derived TFLOPS. The profiler reported 64\% utilization while, by our calculations, we only achieved 49\%. The explanation for this discrepancy is that at $d{=}4096$, the 250~W power budget for our PCIe A100 is exceeded, and the clock is throttled to 1.12 GHz. The implications of this finding are discussed in the conclusion.

\begin{table}[!t]
\caption{Nvidia Nsight Compute profiler results on the \synth dataset with $|D|{=}10^5$. We examine $d\in\{128, 256, 4096\}$ to illustrate how key performance metrics of the brute force TC algorithms (\ouralg and \tedjoinbrute) vary as a function of dimensionality. OOM refers to ``out of shared memory''.}
\footnotesize
    \centering
        \begin{tabular}{l|r|r|r|r|r|r} 
        \hline
           \rowcolor{blue!40!white} Metric &\multicolumn{3}{c|}{\ouralg}& \multicolumn{3}{c}{\tedjoinbrute}\\ \hline
            \rowcolor{blue!40!white}  Dimensionality ($d$)& \cellcolor{blue!40!white}$128$ & \cellcolor{blue!40!white}$256$ &\cellcolor{blue!40!white}$4096$ & \cellcolor{blue!40!white}$128$ & \cellcolor{blue!40!white}$256$&\cellcolor{blue!40!white}$4096$ \\ \hline
             DRAM Throughput (\%) & 1.98 & 3.54 & 16.0 &   0.04 & 0.04 & OOM\\ \hline
             SMEM Throughput (\%) &6.49 & 10.5 & 36.1 &   42.3 & 16.0 & OOM\\ \hline
             Bank Conflicts (\%) & 0.00 & 0.00 & 0.00 & 92.3 & 75.0&OOM \\ \hline
             L2 Hit Rate (\%) & 89.8 & 89.6 &  84.4 &  98.9 & 98.9& OOM\\ \hline
             TC Pipe Utilization FP16-32 (\%) &  10.1& 17.8 & 64.0&   N/A & N/A&OOM \\ \hline
             TC Pipe Utilization FP64 (\%) &  N/A & N/A & N/A & 5.75& 1.99 &OOM\\ \hline
             Clock Speed (GHz) &  1.37 & 1.40 & 1.12 & 1.40 & 1.41 &OOM\\ \hline
        \end{tabular}
    \label{tab:profiler_results}
\end{table}

\subsection{Comparison to the SOTA on Real-World Datasets}
We compare our brute force algorithm, \ouralg, to the SOTA index-supported GPU similarity search algorithms (\mistic, \gdsjoin, and \tedjoinindex) that prune the search to reduce comparing points that are located at sufficiently large distances of each other. This affords the index-supported algorithms a performance advantage over \ouralg as it does not use an index and thus performs $|D|^2$ point comparisons.

\begin{figure*}
    \centering
    \includegraphics[width=\textwidth]{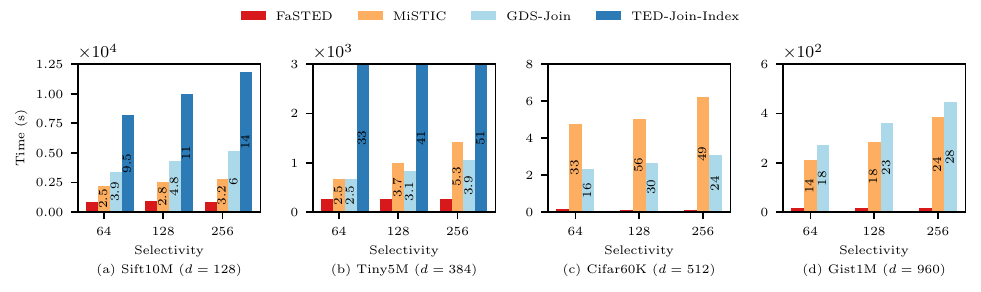}
    \caption{Comparison of \ouralg to the SOTA GPU index-supported similarity search algorithms (\mistic, \gdsjoin, and \tedjoinindex) on four real-world datasets. The total end-to-end response time is plotted, which includes all associated overheads for each method (e.g., index construction and transferring data to/from the GPU). \tedjoinindex exceeds shared memory at $d{=}384$, so it is not shown in panels (c)--(d). The speedup of \ouralg over the reference implementations is reported inside each bar. The standard deviation of the response times for \ouralg in each plot is (a) $\sigma=42$, (b) $\sigma=0.11$, (c) $\sigma=0.12$, and (d) $\sigma=0.054$. In other words, the difference between time trial measurements is negligible on all datasets except for \cifar because the execution time was negligible on that dataset. }
    \label{fig:realworldcomparison}
		\Description{Four plots showing how \ouralg performance compares to three other euclidean distance algorithms. Each plot depicts a different real-world dataset. Each dataset was tested with three different selectivity levels as well. }
\end{figure*}

Figure~\ref{fig:realworldcomparison} plots the response time on the four real-world datasets which span dataset sizes $|D|{=}\text{60,000--10,000,000}$ and data dimensionalities of $d{=}128{-}960$ for the three selectivity values ($S_s{=}64$, $S_m{=}128$, and $S_l{=}256$), which refer to the mean number of neighbors found by each point in each dataset (see Table~\ref{tab:datasets} for details).  Remarkably, we find that \ouralg outperforms all of the index-supported GPU methods and make two observations.

\begin{enumerate}[wide=0pt]
\item On a given dataset, the speedup of \ouralg over the reference implementations increases with increasing selectivity. This is because \ouralg is a  brute force method, and so it computes the distance between all pairs of points, and so the response time of \ouralg is independent of increasing selectivity (or search radius, $\epsilon$). In contrast, the response time of the reference implementations (\mistic, \gdsjoin, and \tedjoinindex) increase with selectivity because the number of distance calculations performed increases with search distance. 
\item The maximum dimensionality on the real-world datasets is $d{=}960$. However, as shown in Section~\ref{sec:eval_dataset_size_dimensionality}, a minimum dimensionality of roughly $d{=}2048$ is needed for \ouralg to reach $\sim$150 TFLOPS. This demonstrates that even in cases where the dimensionality is insufficient to reach $\sim$150 TFLOPS, \ouralg still outperforms the SOTA with a minimum speedup of 2.5$\times$ (over \mistic and \gdsjoin) and a maximum of 51$\times$ over \tedjoinindex.
\end{enumerate}

\subsection{\ouralg: Accuracy of FP16-32}

\ouralg uses FP16-32, so there is greater machine rounding error compared to the FP32 or FP64 reference implementations. We examine the accuracy of \ouralg and configure \gdsjoin to use FP64 values (instead of FP32 used in Figure~\ref{fig:realworldcomparison}) and employ it as the ground truth. We examine two measures of accuracy; the first quantifies the overlap between the points found in the result set for each point in the dataset. The second method compares the difference in the distances computed by the algorithm.

\noindent\textbf{Overlap Between Result Sets:} For each point in the dataset, $p_i\in D$, we compute its neighbors in \ouralg and \gdsjoin, denoted as $N_i^{\ouralg}$ and $N_i^{\gdsjoin}$, respectively. We compare the neighbors for each point in the ground truth set to \ouralg and determine if the sets match. The intersection over the union of the sets for each point yields the overlap between the sets; if the intersection and union of the sets are equal, then we assign an accuracy score of 1.0 for the point, otherwise the score is $<$1.0. Equation~\ref{eqn:accuracy} defines the accuracy score for an entire dataset where a score of 1.0 indicates perfect accuracy:

{
\addtolength{\belowdisplayskip}{-1.0ex}
\addtolength{\belowdisplayshortskip}{-1.0ex}
\addtolength{\abovedisplayskip}{-2.0ex}
\addtolength{\abovedisplayshortskip}{-2.0ex}

\begin{equation}
Accuracy=\frac{1}{|D|}\cdot\sum_{i=1}^{|D|}\frac{\Bigm\lvert N_i^{\ouralg} \cap N_i^{\gdsjoin}\Bigm\lvert}{\Bigm\lvert N_i^{\ouralg} \cup N_i^{\gdsjoin}\Bigm\rvert}.\label{eqn:accuracy}
\end{equation}
}

\begin{table}[!t]
\caption{ The accuracy of \ouralg compared to the FP64 \gdsjoin algorithm using the overlap accuracy metric (Equation 3) across the three selectivity levels for all real-world datasets.}
\footnotesize
    \centering
		\begin{tabular}{p{3.5cm}|r|r|r|r} 
        \hline
             \rowcolor{blue!40!white} & \sift & \tinyds& \cifar & \gist  \\\hline
             $S_{s}{=}64$& 1.0 & 0.99998 &  0.99971 &  0.99999 \\ 
             $S_{m}{=}128$& 1.0 & 0.99997 &  0.99955 &  0.99998 \\ 
             $S_{l}{=}256$& OOM & 0.99996 &  0.99946 &  0.99997 \\ \hline
        \end{tabular}
    \label{tab:accuracy}
\end{table}

\begin{table}[!t]
\caption{The accuracy of \ouralg compared to the FP64 \gdsjoin algorithm using the distance metric with the smallest selectivity level, $S_{s}{=}64$, for all real-world datasets.}
\footnotesize
    \centering
		\begin{tabular}{p{1.6cm}|r|r|r|r} 
        \hline
             \rowcolor{blue!40!white} & \sift & \tinyds& \cifar & \gist  \\\hline
             Mean & $2.6\times 10^{-6}$ & $-1.5\times 10^{-7}$ & $-5.2\times 10^{-7}$  & $-1.6\times 10^{-6}$  \\
             Std. Dev. & $2.4\times 10^{-4}$ & $9.4\times 10^{-6} $ & $3.4\times 10^{-5}$ & $3.7\times 10^{-5}$\\
             \hline
        \end{tabular}
    \label{tab:distance_accuracy}
\end{table}

\begin{figure}
    \centering
    \includegraphics[scale=1.0]{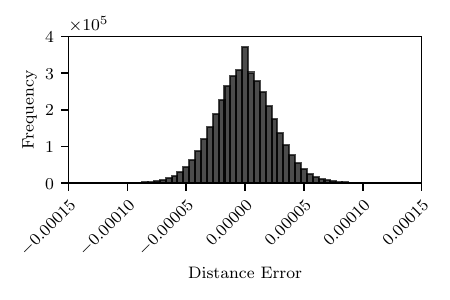}
    \caption{The distribution of the errors in the distance measurements for the \cifar dataset is shown.}
    \label{fig:cifar_distances}
		\Description{Histogram showing the distance error for the Cifar60K dataset. The distribution is centered on 0 error, and has a normal-looking distribution. } 
\end{figure}

\noindent\textbf{Difference Between Computed Distances:} We further our accuracy analysis by examining the difference between the computed distances for each $p_i\in D$ of the points found in the result sets of both \ouralg and \gdsjoin using FP64 (the ground truth). 
The distance error for a single pair of points $p_i$ and $p_j$ is defined as $dist_{i,j}^{\ouralg} - dist_{i,j}^{\gdsjoin}$. We then compute the mean and standard deviation of this error across all pairs of points that appear in both the \ouralg and \gdsjoin result sets.

Table~\ref{tab:accuracy} presents the Overlap Between Result Sets metric across all real-world datasets, where we find that there is a negligible accuracy loss. The maximum accuracy loss occurs on \cifar ($S_l=256$), which still achieves a very high accuracy of 99.946\%.

Table~\ref{tab:distance_accuracy} presents the Difference Between Computed Distances metric for all pairs of points within the $\epsilon$ corresponding to $S_{s}{=}64$. Figure~\ref{fig:cifar_distances} shows the distribution of the errors in the distance calculations for the \cifar dataset (the dataset with the highest error, described above). The distribution in Figure~\ref{fig:cifar_distances} and summary statistics in Table~\ref{tab:distance_accuracy} show that the distances \ouralg computes have no measurable bias and have minimal error. Thus, for similarly distributed datasets that are commensurate with the dynamic range of FP16, \ouralg can compute Euclidean distances without concern for accuracy loss due to machine rounding error. We observed a slight decrease in accuracy with increasing search radius. This detail may be relevant to a reader interested in applications where large distances are of particular importance. In the case of distance similarity searches, and other algorithms that are interested in the local neighborhood of points (e.g., $k$ nearest neighbors), the error in the distance computed is not of consequence.

\section{Discussion \& Conclusions}\label{sec:conclusion}

To our knowledge, we have proposed the first algorithm in the literature that computes Euclidean distances using mixed precision TCs. Similarly, others have successfully used mixed precision arithmetic in other application domains. Cui~\cite{cui2024acceleration} used mixed precision TCs on GPUs for finite element methods, Genome-Wide Association Studies~\cite{ltaief2024toward} employed mixed precision arithmetic on GPU TCs, Geostatistics applications~\cite{cao2022reshaping} and plasma turbulence simulations~\cite{idomura2020acceleration} used mixed precision arithmetic on the A64FX CPU. Many of these papers demonstrate that the loss of accuracy can be recovered, and in some cases, they describe that using FP32 or FP64 may be excessive. Similarly to these other papers, we find that the accuracy loss of our mixed precision computation for distance similarity searches is negligible.

We find that our method scales well with increasing data dimensionality and only requires a small to moderate dataset size to reach our maximum throughput of roughly 150 TFLOPS. We find that compared to the SOTA GPU distance similarity search algorithms that employ an index to prune the search, our approach is significantly faster, despite performing more distance calculations. \ouralg's performance can be primarily attributed to carefully designed optimizations that enable sufficient data reuse, hide memory access latency, and eliminate shared memory bank conflicts.  There are numerous algorithms that employ Euclidean distance calculations, so our algorithm will be of great utility to many application areas beyond similarity searches.

 Despite the datasets that we used not being normalized to the full range of FP16 values, the lowest search accuracy is $99.946\%$, which is an impressive finding. It is likely that scaling the input data could further increase the accuracy of our results, and in the case where a dataset is adversely affected by conversion to FP16, it would mitigate this numerical sensitivity. Future work will investigate this research avenue.

We believe that our maximum performance of $\sim$150 TFLOPS is an underestimate of \ouralg's potential. We observe that when our algorithm executes with sufficiently large dataset sizes and dimensionalities, the clock speed is dynamically reduced because we exceed the 250~W power budget of our PCIe A100. For example, if we were to use an SXM A100 that has a 400~W power budget, we believe that \ouralg would achieve even greater throughput. Thus, the 150~TFLOPS result reported in this paper should be considered the lower bound on the performance of \ouralg when using $|D|{\geq}46416$ and $d{\geq}2048$.

Given that we have significantly utilized the capabilities of modern GPU hardware, future work should focus on efficiency optimizations, such as incorporating an indexing data structure to prune the search, which would further improve performance. However, such an algorithm  will have to be carefully designed to ensure that TC performance is not severely diminished.

While we have reached a performance ceiling for \ouralg, we found that the FP64 TC algorithm in the literature, \tedjoin, is only reaching 6.8\% of peak performance. We hope that the detailed rationales, description, and study of our optimizations will enable them to be applied to redesigning \tedjoin and other GPU-accelerated algorithms in the literature to realize significant performance improvements.

\begin{acks}
This material is based upon work supported by the National Science Foundation under Grant No. 2042155. We thank Benoit Gallet for updating his repository so we could execute experiments with \tedjoin and Brian Donnelly for kindly running experiments with \mistic.
\end{acks}

\bibliographystyle{ACM-Reference-Format}
\bibliography{sample-base}

\end{document}

%% file: textResources.tex
\usepackage{xspace}
\usepackage{amsmath, amsfonts}

\newcommand\kslice{$k$-slice\xspace}
\newcommand\kslices{$k$-slices\xspace}
\newcommand\eightd{$d{=}8$\xspace}
\newcommand\sixtd{$d{=}16$\xspace}
\newcommand\sixfourd{$d{=}64$\xspace}

\newcommand\regf{\text{rf}}
\newcommand\blockf{\text{bf}}
\newcommand\arf{\ensuremath{\bm{A}_{\regf}}}
\newcommand\prf{\ensuremath{\bm{P}_{\regf}}}
\newcommand\qrf{\ensuremath{\bm{Q}_{\regf}}}
\newcommand\pbf{\ensuremath{\bm{P}_{\blockf}}}
\newcommand\qbf{\ensuremath{\bm{Q}_{\blockf}}}